\documentclass{llncs} %[10pt,a4wide]
\usepackage{url,alltt}
\usepackage{amssymb}

\usepackage{epsfig}
\usepackage{wrapfig}
\usepackage{subfigure}
\usepackage{rotating}
\usepackage{moreverb}
\usepackage{fancyvrb}

\def\bd{\begin{description}}
\def\ed{\end{description}}
\def\bc{\begin{center}}
\def\ec{\end{center}}
\def\bq{\begin{quote}}
\def\eq{\end{quote}}
\def\bi{\begin{itemize}}
\def\ei{\end{itemize}}
\def\be{\begin{enumerate}}
\def\ee{\end{enumerate}}
\def\ba{\begin{array}}
\def\ea{\end{array}}

\newcommand{\chr}{{CHR}}
\newcommand{\true}{{\it true}}

\newcommand{\false}{{\it false}}
\newcommand{\simp}{{\; \Leftrightarrow \;}}

\newcommand{\prop}{{\; \Rightarrow\; }}
\newcommand{\com}{\rule{1pt}{7pt}}

\newcommand{\CT}{\ensuremath{\mathcal{CT}}}

\newcommand{\bx}{{\bar x}}

\begin{document}

\title{Why Can't You Behave? %Failure and 
Non-Termination Analysis of Direct Recursive Rules with Constraints} 
\author{%
Thom Fr{\"u}hwirth
}
\institute{%
Ulm University, Germany\\
\email{thom.fruehwirth@uni-ulm.de}
}

\maketitle

\begin{abstract}

This paper is concerned with rule-based programs that go wrong. 
The unwanted behavior of rule applications is non-termination or failure of a computation.
We propose a static program analysis of the non-termination problem for recursion in the Constraint Handling Rules (CHR) language. 

CHR is an advanced concurrent declarative language involving constraint reasoning. 
It has been closely related to many other rule-based approaches, so the results are of a more general interest.
In such languages, non-termination is due to infinite applications of recursive rules.
Failure is due to accumulation of contradicting constraints during the \mbox{computation}. % prevent hyphenation

We give theorems with so-called misbehavior conditions for potential non-termination and failure (as well as definite termination) of linear direct recursive simplification rules.
Logical relationships between the constraints in a recursive rule play a crucial role in this kind of program analysis.
We think that our approach can be extended to other types of recursion and to a more general class of rules. 
Therefore this paper can serve as a basic reference and a starting point for further research.
\end{abstract}

This is the full version of the paper. 
The final publication \cite{nonterm2016} is available at Springer via \url{http://dx.doi.org/10.1007/978-3-319-42019-6_14}.

\section{Introduction}

It is well known that termination is undecidable for Turing-complete programming languages. Thus, there is a long tradition in research on program analysis methods, static and dynamic, to tame the problem by semi-automatic or approximative approaches.

In this work we are interested in characterizing non-terminating computations. We do so 
in the context of the programming language Constraint Handling Rules 
(CHR)~\cite{fru_chr_book_2009,fruhwirth2015constraint,chrwebsite}.
As in other rule-based languages, termination is only an issue if recursion is involved.
We are hopeful that our results could be transferred to other rule-based programming languages as well, since CHR can directly embed many rule-based languages and formalisms (e.g. Chapter 6 in~\cite{fru_chr_book_2009}).

We propose conditions for misbehavior, i.e. a static program analysis of a recursive rule that tells us if a given goal (or a set of goals) may not terminate or lead to failure (unsatisfiable constraints).
The following program serves as a first overview of the characteristic features of CHR 
for those not familiar with the language. 
In CHR, we use a first-order logic syntax.
Predicates will be called constraints.
Goals and states are synonyms here, they are conjunctions of constraints.
In this paper, numbers are expressed in successor term notation.
The following example will be further elaborated in this paper.
\begin{example}\label{double-intro}
{\rm 
Consider a recursive user-defined constraint $\mathit{double}$ that doubles the natural number in the first argument and produces the resulting number in the second argument:

\medskip
$\begin{array}{l}
\mathit{double}(X,Y) \simp X{=}0 \; \com \; Y{=}0.\\
\mathit{double}(X,Y) \simp X{=}s(X1) \; \com \; Y{=}s(s(Y1)) \land \mathit{double}(X1,Y1).
\end{array}$
\medskip

The first rule (for the base case) says that if $X$ is syntactically equivalent to $0$, 
then the result $Y$ is also zero. 
The syntactic equality constraint $X{=}0$ is a {\em guard}, a precondition on the applicability of the rule.
It serves as a test.
The rule is only applied if this condition holds in the current context, i.e. state.
On the other hand, $Y{=}0$ is a constraint that is asserted once the rule is applied.
The recursive rule says that if $X$ is the successor of some number $X1$, then $Y$ is 
the successor of the successor of some number $Y1$, and $X1$ doubled gives $Y1$. 

To the goal $\mathit{double}(X,Y)$ no rule is applicable.
To the goal $\mathit{double}(X,Y) \land X{=}0$ the first rule is applicable, resulting in the state $X{=}0 \land Y{=}0$.
To the goal $\mathit{double}(X,Y) \land X{=}0 \land Y{=}s(B)$, the first rule is also applicable, but the resulting contradiction $Y{=}s(B) \land Y{=}0$ means failure due to these unsatisfiable equality constraints.

In logical languages like CHR, variables cannot be overwritten, but they can be without value (unbound). 
For example, if $X$ is $s(A)$, where $A$ is unbound, then $X$ will satisfy the guard, and $Y$ will be equated to $s(s(B)))$, where $B$ is some newly introduced variable, and the CHR constraint $\mathit{double}(A,B)$ will be added to the state. Since $A$ is unbound, the guard for the recursive goal $\mathit{double}$ does not (yet) hold.
If the variable later becomes (partially) bound in a syntactic equality, the
computation of $\mathit{double}$ may resume.

There is a simple example for a infinite computation with $\mathit{double}$.
The goal $\mathit{double}(X,Y) \land X{=}s(X1) \land X{=}Y$  does not terminate.
The application of the recursive rule leads to the state 
$X{=}s(X1) \land X{=}Y \land Y{=}s(s(Y1)) \land \mathit{double}(X1,Y1)$.
Since $X{=}Y$, we have that $X1{=}s(Y1)$.
Thus the computation can proceed with another recursive rule application and so on ad infinitum. The successors that are produced for $Y$ in the second argument will also become successors for $X$ in the first argument, because of $X{=}Y$. Thus the guard of the recursive goal always holds.

The goal $\mathit{double}(X,Y) \land X{=}s(X1) \land X{\geq}Y$ does not terminate either.
Our main theorem will allow to detect this non-termination because a misbehavior condition holds. 
Basically, the guard and the body of the recursive rule, $X{=}s(X1) \land Y{=}s(s(Y1))$, together with the added constraint $X{\geq}Y$ implies the guard of the recursive goal $X1{=}s(X1')$.
Another theorem will tell us that the more stricter constraint $X{=}Y$ will therefore inherit the misbehavior.

}
\end{example}

\noindent{\bf Related Work.}
Non-termination analysis has been considered for term rewriting systems \cite{payet08,endrullis2015proving},
logic programming languages \cite{deschreye,kifer,mesnardcp,fruhwirth2015devil}, %
and imperative languages \cite{popl2008,Brockschmidt,mesnardjava,le2015termination}.

The works on (constraint) logic programming are based on finding loops in abstracted partial derivation trees. 
In our restricted case of linear direct recursion it is sufficient to consider the recursive rule and no abstraction is necessary.
However, there is also a difference to our approach: mode information about the arguments is essential in analysing logic programs. 
A similar type of information is also needed for non-termination of constraint logic programs \cite{mesnardcp}. 
It gives raise to so-called filters for abstracting states of the computation. 
In CHR, this information is already implicitly encoded in the distinction between guard and body built-in constraints.

In \cite{fruhwirth2015devil} a simple program transformation for recursive rules in CHR was introduced that produces one or more adversary rules. 
When the rules are executed together, a non-terminating computation may arise. It was shown
that any non-terminating computation of the original rule contains this witness computation. 
Based on the adversary rules, a preliminary condition for non-termination was proposed. This condition only refers to the witness computation that starts from a particular state, it can be considered as one particular special case of the misbehavior conditions we give here.

\medskip
\noindent{\bf Overview of the paper.}
In the next section we define syntax and operational semantics for CHR simplification rules.
Section 3 gives a first basic theorem for non-termination or failure of a specific (the most general) goal for a given linear direct recursive rule.
Section 4 gives our main condition for misbehavior of a recursive rule in a generalised theorem. 
Another theorem shows that any goal that contains a misbehaved goal will also be misbehaved.
We end the paper with conclusions and directions for future work.

\section{Preliminaries}\label{sec:chr}

In this section we give a restricted overview of syntax and semantics for Constraint
Handling Rules (CHR)~\cite{fru_chr_book_2009}, cut down to what is essential for this paper (namely simplification rules).
We assume basic familiarity with first-order predicate logic and state transition systems.
Readers familiar with CHR can skip this section.
CHR is a committed-choice language, i.e. there is no backtracking in the rule applications. 
CHR is a concurrent language, i.e. we may apply rules in parallel.

\subsection{Abstract Syntax of \chr}

Constraints are distinguished predicates of first-order predicate logic.
We distinguish between 
two different kinds of constraints: {\em built-in (or: pre-defined) constraints} 
which are handled by a given constraint solver,
and
{\em user-defined (or: CHR) constraints} 
which are defined by the rules in a CHR program.
A {\em \chr\ program} is a finite set of rules.  
There are two basic kinds of rules in CHR:

\begin{center}
\begin{tabular}{ll}
{\em Simplification rule:} & {\it r} $: \; H \simp C \; \com \; B,$ \\
{\em Propagation rule:} & {\it r} $: \; H \prop C \; \com \; B,$ \\
\end{tabular}
\end{center}

\noindent where {\em r:} is an optional, unique identifier of a rule, the {\em head}  
$H$ is a non-empty conjunction of user-defined constraints, the
{\em guard} $C$ is a conjunction of built-in constraints, and the {\em body} $B$
is a goal. A {\em goal} is a conjunction of built-in and CHR
constraints.
An {\em empty} guard expression $\true \; \com$ can be omitted from a rule.

In this paper, we are only concerned with a simple class of simplification rules, 
so propagation rules will be ignored from now on.

\subsection{Abstract Operational Semantics of \chr}\label{sec:chr:semantics}

Computations in CHR are sequences of rule applications.
The operational semantics of CHR is given by the state transition system. %!!! in Figure~\ref{trans}.
(Concurrency is not made explicit in the semantics given, 
since it is independent of the results of this paper.)
{\em States} are goals.
Let $\CT$ be a constraint theory for the built-in constraints,  
including the trivial $\true$ and $\false$ as well as syntactical equality $=$ over finite terms.
For a goal $G$, the notation $G_{bi}$ denotes the built-in constraints of $G$
and $G_{ud}$ denotes the user-defined constraints of $G$.

In the transition system, all single upper-case letters are meta-variables that stand for goals.
Let the variables in a disjoint variant of a rule be denoted by $\bar x$.
A {\em disjoint (or: fresh) variant} of an expression is obtained by uniformly replacing its
variables by different, new (fresh) variables. 
A {\em variable renaming} is a bijective function over variables.

\begin{center}
  {\bf Simplify State Transition of CHR}\\[5pt] %!!!
  \begin{tabular}{l@{\quad}l}
    If   & %\parbox[t]{.8\textwidth}{\raggedright
      $(r: \; H \Leftrightarrow C \; \com \; B)$ is a disjoint variant of
      a rule in the program\\ %!- with variables $\bar x$\\  %!! program $P$
    and  & $\CT \models \exists (G_{bi}) \land %!
			\forall (G_{bi} \to \exists \bar x (H{=}H_S \land C))$\\
    then & $({H_S} \land G) \mapsto_r (B \land  G
        \land H{=}H_S \land C)$
    %$\begin{array}[c]{cc}
  \end{tabular}

\label{trans}
\end{center}

Starting with a given initial state, 
CHR rules are applied exhaustively, until a fixed-point is reached.
A simplification rule 
$H \simp C \; \com \; B$
that is applied
removes the user-defined constraints matching $H$ and replaces them by $B$ provided the guard
$C$ holds.  
Note that built-in constraints in a computation are accumulated, i.e. they are added but never removed, while user-defined constraints can be added as well as removed. 
The built-in constraints allow execution in the abstract without the need to know values for variables, 
just their relationships are expressed as constraints.

A rule is {\em applicable}, if 
its head constraints are matched by constraints in the current goal
one-by-one and if, under this matching, 
the guard of the rule is logically implied by the built-in constraints in the goal,
provided they are satisfiable.
Any one of the applicable rules can
be applied in a transition, and the application cannot be undone, it is committed-choice.
An expression of the form 
$\CT \models \exists (G_{bi}) \land %! 
			\forall (G_{bi} \to \exists \bar x (H{=}H_S \land C))$
is called {\em applicability condition}.
We may drop $\CT \models$ for convenience later on.
We use ${H}{=}{H_S}$ by abuse of notation, since the arguments of this syntactic equality are conjunctions of user-defined constraints. This expression means to pairwise equate the user-defined constraints on the left and right hand side and then to pairwise equate their arguments, which are terms.

In a transition (or: {\em computation step}) 
$S \mapsto_r T$, $S$ is called {\em source state} and $T$ is called {\em target state}.
When it is clear from the context, we will drop the reference to the rule $r$.
A {\em computation} of a goal $G$ in a program $P$
is a connected sequence
$S_i \mapsto S_{i+1}$ beginning with
the {\em initial state} $S_0$ that is $G$ 
and ending in a {\em final state} or the sequence is {\em non-terminating (or: diverging)}.
The notation ${\mapsto}^*$ denotes the reflexive and transitive closure of $\mapsto$.

A goal (state) is {\em satisfiable (consistent)} if its built-in constraints are satisfiable.
A state with unsatisfiable (inconsistent) built-in constraints is called a {\em failed} state.
A computation of a goal is failed if it ends in a failed state.
If a computation of a goal is failed (non-terminaing), we may also say that the goal is failed (non-terminating).

Two states $S_1 = (S_{1bi} \land S_{1ud})$ and $S_2 = (S_{2bi} \land S_{2ud})$ are 
{\em equivalent} as defined in~\cite{betz_raiser_fru_execution_model_iclp10}, 
written $S_1 \equiv S_2$, if and only if
$$	\CT \models 
        \forall (S_{1bi} \rightarrow \exists \bar y ((S_{1ud} = S_{2ud}) \land S_{2bi}))
	\land %\\
	%\indent \indent 
         \forall (S_{2bi} \rightarrow \exists \bar x ((S_{1ud} = S_{2ud}) \land S_{1bi}))
$$
\noindent with $\bar x$ those variables that only occur in $S_1$ and $\bar y$ those variables that only occur in $S_2$.
A goal (or state) $S$ is (strictly) %!!
{\em contained ({\rm or:} included)} in a goal $T$ (or: less specific than $T$)
 if and only if 
there exists a (non-empty) goal $G$ such that $(S \land G) \equiv T$.

Note that this notion of state equivalence is stricter than logical equivalence since it 
it considers multiple occurrences of user-defined constraints to be different as in a multiset.
For this reason, state equivalence is defined by two symmetric implications and syntactically equates the two states.

\section{A Basic Misbehavior Condition for Non-Termination} % and Failure}

In this paper we are concerned with linear direct recursion, expressed by simplification rules of the form
\[r: H \simp C \; \com \; B_{bi} \land B_{ud},\]
where $H$ and $B_{ud}$ are atomic user-defined constraints for the same predicate symbol
and where $C$ and $B_{bi}$ are built-in constraints.

To introduce our appropach, we will start with a theorem about a condition for non-termination 
that only applies to a specific initial goal.
It is not just any goal, however. It is of the form $H \land C$, i.e. it consists of the head and guard of the given recursive rule. Such a goal is the {\em most general state} to which the rule is applicable. This is easy to see, since removing $H$ or replacing $C$ by more general, weaker built-in constraints would invalidate the rule application condition of the operational semantics of CHR.

The theorem below already reflects the structure of the upcoming main theorem. Certain goals for a given recursive rule are non-terminating or failing if a certain implication between the built-in constraints of the guard and body of the rule holds. 
Our theorems provide an analysis that does not distinguish between non-termination and (termination by) failure of goal. This is justifiable, since in both cases the computation goes wrong. We therefore refer to the conditions in the theorems as {\em misbehavior} conditions.

The misbehavior condition we give is typically decidable (depending on the decidability of the underlying theory for the built-in constraints, of course). Since termination (the halting problem) is undecidable for Turing-complete programming languages, we cannot expect a sufficient and necessary condition in general. A sufficient condition suffices. 
Interestingly, for the most general goal $H \land C$ of a rule, we can give a condition that clearly separates termination from non-termination, but is agnostic to failure.
This is what the first theorem is about (and it sets the stage for a more general theorem).

\begin{theorem}\label{nontermin1}
{\rm
Given a recursive rule 
\[r: H \simp C \; \com \; B_{bi} \land B_{ud},\]
and its disjoint variant with variables $\bx$
\[r: H' \simp C' \; \com \; B'_{bi} \land B'_{ud},\]
then the {\em basic misbehavior condition}
\[\CT \models \exists (C \land B_{bi})
 \ \land\] 
\[ \ \ \  \ \ \ \forall ((C \land B_{bi}) \to \exists \bx ({B_{ud}}{=}{H'} \land C')).\]
implies non-termination or failure of the goal 
\[H \land C\]
through rule $r$.

\medskip
If the basic misbehavior condition does not hold, then the computation of the goal 
\[H \land C\]
through rule $r$ terminates.

\medskip
\noindent {\bf Proof.} 
The proof can be found in the appendix of the full version of this paper that is available online via the authors homepage. %!!!!
It is based on the proof of a more general theorem that will be stated in the next section.
\qed}\end{theorem}

Note that while non-termination requires the basic condition to hold, 
failure of the goal may occur whether the condition holds or not.
So the condition is necessary for non-termination of the goal $H \land C$, 
but does not make a statement about failure.
Thus the condition is not sufficient for non-termination, 
but it is sufficient for misbehavior (non-termination or failure).
Still it is remarkable that we can give a converse of this misbehavior condition. 
This will not be the case any more for the general theorem.

\medskip
We now look at some examples to see applications of this first theorem.
\begin{example}\label{number}
{\rm 
Here is a simple recursive rule that goes through the successors that define a natural number:
\[\mathit{number(X)} \simp X{=}s(Y) \; \com \; \mathit{number(Y)}.\]
Note that there are no built-in constraints in the body of the rule.

The basic misbehavior condition amounts to
\[\CT \models \exists {XY} (X{=}s(Y))
 \ \land\] 
\[ \ \ \  \ \ \ \forall {XY} ((X{=}s(Y)) \to \exists {X'Y'}({\mathit{number(Y)}}{=}{\mathit{number(X')}} \land X'{=}s(Y'))).\]
The first, existential part of the condition holds, while the implication in the second part does not. 
It is not the case that for all $Y$, $Y$ is equivalent to some $X'$ that in turn is equivalent to $s(Y')$. 
For example, $Y$ may be $0$.
Thus the goal $number(X) \land X{=}s(Y)$ will terminate.
Actually it will lead to the state $X{=}s(Y) \land \mathit{number(Y)}$.

Now consider a variant of the above rule that enforces the constraint that a variable must be a successor term:
\[\mathit{number(X)} \simp X{=}s(Y) \land \mathit{number(Y)}.\]
Note that there are no built-in constraints in the guard of the rule, so the guard has been dropped.
The basic misbehavior condition amounts to
\[\CT \models \exists {XY} (X{=}s(Y))
 \ \land\] 
\[ \ \ \  \ \ \ \forall {XY} ((X{=}s(Y)) \to \exists {X'Y'}({\mathit{number(Y)}}{=}{\mathit{number(X')))}}.\]
This condition holds, there fore the goal $number(X) \land X{=}s(Y)$ will not terminate or lead to failure.
Actually, it will not terminate, producing a longer and longer nested term of successsors.

Next consider a variant of the first rule where the position of the variables $X$ and $Y$ is interchanged in the guard constraint:
\[\mathit{number(X)} \simp Y{=}s(X) \; \com \; \mathit{number(Y)}.\]
The basic misbehavior condition amounts to
\[\CT \models \exists {XY} (Y{=}s(X))
 \ \land\] 
\[ \ \ \  \ \ \ \forall {XY} ((Y{=}s(X)) \to \exists {X'Y'}({\mathit{number(Y)}}{=}{\mathit{number(X')}} \land Y'{=}s(X'))).\]
The condition holds, since for all $Y$ that are equivalent to $X'$, there exists a $Y'$ such that $Y'{=}s(X')$.
And indeed, the goal $number(X) \land Y{=}s(X)$ will not terminate.
}
\end{example}

\begin{example}\label{double1}
{\rm 
Consider the recursive rule for the constraint $\mathit{double}$ from 
Example~\ref{double-intro} of the introduction section:
\[\mathit{double(X,Y)} \simp X{=}s(X1) \; \com \; Y{=}s(s(Y1)) \land \mathit{double(X1,Y1)}.\]
The implication of the basic misbehavior condition is
\[\forall ((X{=}s(X1) \land Y{=}s(s(Y1))) \to \exists ({\mathit{double(X1,Y1)}}{=}{\mathit{double(X',Y')}} \land X'{=}s(X1'))).\]
It does not hold. Actually, the goal $\mathit{double(X,Y)} \land X{=}s(X1)$ is terminating and does not fail.
The rule can be applied once.
}
\end{example}

\begin{example}\label{notrecursive1}
{\rm 
Consider the following rule with empty guard and $X{>}Y$ in its body
\[p(X,Y) \simp X{>}Y \land p(Y,X).\]
The implication of the misbehavior condition then is
\[ \ \ \  \ \ \ \forall XY ((X{>}Y) \to \exists X'Y' ({p(Y,X)}{=}{p(X',Y')}).\]
Clearly, the basic misbehavior condition holds. 
Actually, the goal $p(X,Y)$ will fail at the second recursive step, since the recursive call exchanges the two arguments of $p$ but $X{>}Y$ and $Y{>}X$ contradict each other.
}
\end{example}

\begin{example}\label{prime} %! user tabular or array format layout
{\rm 
Let {\em odd} and {\em prime} be built-in constraints.   %! constraints odd and prime used only here
Consider the following recursive rule
\[c(X) \simp \mathit{odd}(X)  \; \com \;  c(s(s(X))),\]
The misbehavior condition amounts to
\[\exists X \mathit{odd}(X)  \land \forall X (\mathit{odd}(X) \to \exists X' (c(s(s(X))){=}c(X') \land \mathit{odd}(X'))).\]
Since the successor of the successor of an odd number is always odd, the condition holds.
Indeed, the goal $c(X) \land \mathit{odd}(X)$ is non-terminating.

Now consider a variation of the above rule
\[c(X) \simp \mathit{prime}(X)  \; \com \;  c(s(s(X))).\]
The condition amounts to
\[\exists X \mathit{prime}(X)  \land \forall X (\mathit{prime}(X) \to \exists X' (c(s(s(X))){=}c(X') \land \mathit{prime}(X'))).\]
Since the successor of the successor of a prime number may not be prime, the condition does not hold. 
Thus the goal $c(X) \land \mathit{prime}(X)$ terminates.
It does so after one recursive step.
(It will terminate for any given number $X$ in at most two recursive steps: 
one of every three sequential even or odd natural numbers is a multiple of three, and hence not prime.)
} %rm
\end{example}

\section{The Main Misbehavior Condition}

We are going to state a generalization of Theorem~\ref{nontermin1}.
It is easy to see from the CHR operational semantics and its applicability condition that any state to which a given rule is applicable must contain its head and guard. All such states are therefore equivalent to a state of the form $H \land G \land Q$, where $Q$ is an arbitrary constraint. 

To generalise our initial theorem, we could simply add $Q$ to the premise of the implication in the basic misbehavior condition. This is, however, not sufficient to guarantee non-termination or failure. As it turns out, we also have to add an appropriate variant of $Q$ to the conclusion of the implication. This ensures that the appropriate variant of $Q$ holds at each recursive step.
This will be our main misbehavior theorem.

We will then show in another theorem that any state that contains $H \land G \land Q$ which misbehaves is also doomed to misbehave. So both theorems together typically cover an infinite set of states that do not terminate or fail.

\subsection{Lemmata}

For the proof of the upcoming main theorem, we will need the following lemmata. 

\begin{lemma}\label{devils1} %\cite{fruhwirth2015devil}
{\rm
Given goal $C$ consisting of built-in constraints only and a goal $H$ consisting of user-defined constraints only.
Let the pairs $(H,C)$ with variables $x$ and $(H',C')$ with variables $y$ be disjoint variants.
Then the applicability condition
\[\CT \models \forall \bar x (C \to \exists \bar y (H'{=}H \land C'))\]
trivially holds.

The proof follows from the tautologies of first order predicate logic and can be found in \cite{fruhwirth2015devil}.
} %rm
\end{lemma}

The following lemma states that built-in constraints are accumulated over the course of a computation.
\begin{lemma}\label{addbuiltins}
{\rm
If there is a computation from state $G$ to state $G'$, 
then the built-in constraints of $G'$ imply those of $G$.
\[\mbox{If } G%_\glob 
\mapsto^* G'%_\glob
  \mbox{ then }(G_{bi} \leftarrow G'_{bi}).\]

\medskip
\noindent {\bf Proof.} 
This claim can be shown by comparing the constraints in the source and target states of a computation step according to the operational semantics of CHR,
$({H_S} \land G) \mapsto (B \land  G \land H{=}H_S \land C)$.
The source state has built-in constraints $G_{bi}$, 
the target state has built-in constraints $B_{bi} \land  G_{bi} \land H{=}H_S \land C$.

Since the implication holds for a single transition $\mapsto$, 
it also holds for an arbitrary number of transitions $\mapsto^*$, 
since logical implication $\leftarrow$ is reflexive and transitive.
\qed}\end{lemma}
In other words, if the source state contains built-in constraints, the target state contains them as well.

The next lemma
states an important monotonicity property of CHR. 
It is a variation of the classic CHR monotonicity Lemma (Sect. 4.2. in \cite{fru_chr_book_2009}).
It can be proven from the operational semantics of CHR 
using Lemma~\ref{addbuiltins} and induction over the computation steps.
\begin{lemma}\label{monotonicity}
{\rm
{\em (CHR monotonicity)}
If a rule $r$ is applicable to a state, 
it is also applicable to the state when constraints have been added, 
as long as this state is not failed.
\[\mbox{If } G%_\glob 
\mapsto_r G'%_\glob
  \mbox{ then }(G \land H)%_{\glob'} 
\mapsto_r (G' \land H)%_{\glob'}
,\]
provided $G \land H$ is satisfiable.

A computation can be repeated in any larger
context, i.e.\ with states in which constraints have been
added. 
\[\mbox{If } G%_\glob 
\mapsto^* G'%_\glob
  \mbox{ then }(G \land H)%_{\glob'} 
\mapsto^* (G' \land H)%_{\glob'}
\textrm{ or the computation diverges}.\] %!!!
}
\end{lemma}
Note that $G'$ may be satisfiable while $G' \land H$ may be a failed state.

\subsection{Main Misbehavior Theorem}

We are now ready to state the main theorem of the paper.
\begin{theorem}\label{nontermin2}
{\rm
Let $Q$ be a built-in constraint.
Given $Q$ and a recursive rule 
\[Q, \ \ \ r: H \simp C \; \com \; B_{bi} \land B_{ud},\]
and their disjoint variant with variables $\bx$
\[Q', \ \ \ r: H' \simp C' \; \com \; B'_{bi} \land B'_{ud},\]
Then the {\em general misbehavior condition}
\[\CT \models \exists (Q \land C \land B_{bi})
 \ \land\]
\[\forall ((Q \land C \land B_{bi}) \to \exists \bx ({B_{ud}}{=}{H'} \land Q' \land C')).\]
implies non-termination or failure of the computation of the goal 
\[H \land C \land Q\]
through rule $r$.

\medskip
\noindent {\bf Proof.} 
We prove the claim by induction over the computation steps.

\medskip
{\bf Base Case.}
The claim is that the goal $H \land C \land Q$ either is failed or 
there exists a computation step by applying the recursive rule $r$.
We show that there is always such a computation step possible 
(and that the resulting state is not failed).

According to the abstract operational semantics of CHR, this computation step must be of the form:
  %\begin{tabular}{l@{\quad}l}
      $$(H \land C \land Q) \mapsto_r (B'_{bi} \land B'_{ud} \land  C \land Q
        \land H'{=}H \land C')$$
    $${\rm if }  \ \ \CT \models  \exists (C \land Q) \land 
			\forall (C \land Q \to \exists (H'{=}H \land C'))$$
   %\end{tabular}

We have to show that the applicability condition holds, so that we can apply the recursive rule. 
By the first, existential part of the general misbehavior condition 
we know that $\exists (Q \land C \land B_{bi})$ is satisfiable.
Since this conjunction logically implies $\exists (C \land Q)$, we know that the source state 
$\exists (H \land C \land Q)$ is satisfiable, too.
By Lemma~\ref{devils1} we know that $\forall (C \to \exists (H'{=}H \land C'))$ trivially holds. 
So $\forall (C \land Q \to \exists (H'{=}H \land C'))$ holds as well.
Thus the applicability conditions holds and the recursive rule $r$ is applicable.

The resulting target state of the transition is $(B'_{bi} \land B'_{ud} \land C \land Q
        \land H'{=}H \land C')$. 
$B'_{ud}$ is a user-defined constraint and thus can be ignored for determining the satisfiability of the state.
We already know from the applicability condition that $\exists (C \land Q
        \land H'{=}H \land C')$.
By Lemma~\ref{devils1} we know that $\forall (C \land B_{bi} \to \exists (H'{=}H \land C' \land B'_{bi}))$ trivially holds.
By the first part of the misbehavior condition we know that $\exists (Q \land C \land B_{bi})$ is satisfiable.
Thus $(B'_{bi} \land  C \land Q
        \land H'{=}H \land C')$
must also be satisfiable.
Thus the target state is satisfiable.

\medskip
{\bf Inductive Step.}
We have to show that given a state where the recursive rule has been applied, either the recursive rule is applicable again or the state is failed.

We assume such states are of the form 
$(G \land B_{bi} \land B_{ud} \land C \land Q)$, where $G$ is an arbitrary constraint.
This form holds for the target state of the base case.

Now consider a source state of the desired form. If it is failed, we are done.
If it is not failed, we show that the following computation step is possible with the recursive rule:
     $$(G \land B_{bi} \land B_{ud} \land C \land Q) \mapsto_r (G \land B_{bi} \land B'_{bi} \land B'_{ud} \land  C \land Q
        \land H'{=}B_{ud} \land C')$$
    $${\rm if }  \ \ \CT \models  \exists (G_{bi} \land B_{bi} \land C \land Q) \land 
		\forall (G_{bi} \land B_{bi} \land C \land Q \to \exists (H'{=}B_{ud} \land C'))$$
For the proof of applicability of the recursive rule we reuse the one for the base case. Instead of $H'$, we have now $B_{ud}$, and there are additional constraints $G \land B_{bi}$ in the source state. 
By monotonicity of CHR (Lemma~\ref{monotonicity}), we know that if a rule is applicable to a state, it is also applicable to the state when constraints have been added, as long as this state is not failed. 
Thus the additional constraints $G \land B_{bi}$ cannot inhibit the applicability of the rule, since the state is not failed.

We still have to show that the target state is of the required form. But $Q'$ seems to be missing from it.
The implication of the misbehavior condition in the theorem is
\[\forall ((Q \land C \land B_{bi}) \to \exists ({B_{ud}}{=}{H'} \land Q' \land C')).\] 
Therefore, since the target state contains $(Q \land C \land B_{bi})$, 
it also contains $({B_{ud}}{=}{H'} \land Q' \land C')$.
Thus the target state is equivalent to $(G' \land B'_{bi} \land B'_{ud} \land C' \land Q')$, when we let $G'$ be
$(G \land B_{bi} \land C \land Q \land H'{=}B_{ud})$.

So the target state is also of the required form. 
\qed}\end{theorem}
Theorem~\ref{nontermin2} states an implication between the general misbehavior condition and failing or non-terminating goals. The condition is sufficient for misbehavior, but not necessary. 
As we will see, due to the next theorem, the converse does not hold (unlike Theorem~\ref{nontermin1}).

\medskip
We continue with some examples, old and new, for the application of the main misbehavior theorem.
\begin{example}\label{prime1} %! user tabular or array format layout
{\rm 
Consider a variation of the recursive rule from Example~\ref{prime} with the opposite guard:
\[c(X) \simp \mathit{notprime}(X)  \; \com \;  c(s(s(X))).\]
The basic misbehavior condition amounts to
\[\exists X \mathit{notprime}(X)  \land \forall X (\mathit{notprime}(X) \to \exists X' (c(s(s(X))){=}c(X') \land \mathit{notprime}(X'))).\]
Since the successor of the successor of a non-prime may be prime, the condition does not hold. 
By Theorem~\ref{nontermin1}, the goal $c(X) \land \mathit{notprime}(X)$ thus terminates.

Let $Q$ be $\mathit{odd}(X)$. The implication of the general misbehavior condition is
\[\forall X (\mathit{odd}(X) \land \mathit{notprime}(X) \to \exists X' (c(s(s(X))){=}c(X') \land \mathit{odd}(X') \land \mathit{notprime}(X'))).\]
Again, it does not hold.
The status of non-termination is undecided by Theorem~\ref{nontermin2}.
(Actually, there is no infinite sequence of odd numbers that does not contain a prime, therefore any computation containing $c(X) \land \mathit{odd}(X)$ will terminate.)

Now let $Q$ be $\mathit{even}(X) \land X{=}s(s(s(Y)))$. This time the condition holds, since any sequence of even numbers greater or equal to three (since $X{=}s(s(s(Y)))$) does not contain a prime number. The corresponding goal $c(X) \land \mathit{even}(X) \land X{=}s(s(s(Y)))$ is non-terminating.
(So $c(X)$ terminates for odd numbers but does not terminate for even numbers greater than two.)

} %rm
\end{example}

The following example exhibits a non-terminating computation for a list concatentation constraint.
\begin{example}\label{append}
{\rm 
Let $cons$ and $nil$ denote function symbols to build lists.
Then we can define the concatentation of two lists $L1$ and $L2$ resulting in a third list $L3$:

$$append(L1,L2,L3) \simp L1{=}nil \; \com \; L2{=}L3.$$
$$append(L1,L2,L3) \simp L1{=}cons(X,L1') \; \com \; $$
$$ \hspace*{4cm} L2{=}L2' \land L3{=}cons(X,L3') \land append(L1',L2',L3').$$  %\indent

The implication of the basic misbehavior condition is
$$\forall %{L1 L1' L2 L2' L3 L3'} 
(L1{=}cons(X,L1') \land L2{=}L2' \land L3{=}cons(X,L3') \to$$
$$\exists %{X'L1''L1'''L2''L3''} 
(append(L1',L2',L3'){=}append(L1'',L2'',L3'') \land L1''{=}cons(X',L1''')))$$
This formula does not hold, because the premise of the implication does not constrain $L1''$ (which is equivalent to $L1'$) to be a $cons$ term as required by the conclusion.

Regarding the general misbehavior condition, let $Q$ be $L1'{=}L3$. 
Then the implication of the general condition amounts to
$$\forall (L1'{=}L3 \land L1{=}cons(X,L1') \land L2{=}L2' \land L3{=}cons(X,L3') \to $$
$$\exists 
(append(L1',L2',L3'){=}append(L1'',L2'',L3'') \land L1'''{=}L3'' \land L1''{=}cons(X',L1''')))$$
This formula does hold, because $L1'{=}L3$ and $L3{=}cons(X,L3')$ in the premise
implies $L1'{=}L1'' \land L1'''{=}L3'' \land L1''{=}cons(X',L1''')$ in the conclusion, 
as we can choose $X'{=}X$ and $L3''{=}L3'$.
Indeed, the computation for the goal 
$$append(L1,L2,L3) \land L1{=}cons(X,L1') \land L1'{=}L3$$ 
is non-terminating, producing longer and longer lists.
} %rm
\end{example}

\subsection{Containment Theorem}  

Theorem~\ref{nontermin2} only gives us a particular goal that is non-terminating or fails. 
By the following theorem we can apply the theorem to any goal that contains that particular goal.
Usually, there are infinitely many such goals. 
The proof directly follows from the previous theorem and the monotonicity property of CHR.
\begin{theorem}\label{nontermin3}
{\rm
Any goal 
$$H \land C \land Q \land G$$ 
with arbitrary constraint $G$ ,
where the general misbehavior condition according to Theorem~\ref{nontermin2},
holds for $H \land C \land Q$,
will either not terminate or fail.

\medskip
\noindent {\bf Proof.} 
We prove the claim by induction.

{\bf Base Case.} The state $H \land C \land Q \land G$ is either failed or not.
In the first case we are done.
In the second case, the recursive rule is applicable to the state.
Because by monotonicity of CHR (Lemma~\ref{monotonicity}), we know that if a rule is applicable to a state, it is also applicable to the state when constraints have been added, as long as this state is not failed. 

{\bf Induction Step.} 
The same reasoning holds for all subsequent states in the computation: If we have a state, it is either failed or the recursive rule is applicable to it by monotonicity.

Thus the computation of a goal $H \land C \land Q \land G$ either fails or diverges.
\qed}\end{theorem}

The following example introduces some specific goals for a non-terminating computation.
\begin{example}\label{notrecursive2}
{\rm 
Consider a variant of the rule of Example~\ref{notrecursive1} with empty guard and $X{\geq}Y$ in its body
\[p(X,Y) \simp X{\geq}Y \land p(Y,X).\]
Let $Q$ be $\true$. The implication of the misbehavior condition then is
\[ \ \ \  \ \ \ \forall XY ((X{\geq}Y) \to \exists X'Y' ({p(Y,X)}{=}{p(X',Y')}).\]
This condition holds. 
So any computation for a goal consisting of $p(X,Y)$ and arbitrary built-in constraints 
either fails or is non-terminating.
The computation for the goal
$p(X,Y)$ is non-terminating. 
So is the more specific goal $p(X,Y) \land X{=}Y$. 
The more specific goal $p(X,Y) \land X{<}Y$ fails.
So do the goals with the built-in constraints $X{>}Y$ and $X{\neq}Y$.
The goal $p(X,Y) \land p(Y,X)$ is non-terminating as well, producing the constraint $X{=}Y$.
}
\end{example}

Note that while $Q$ satisfies the misbehavior condition, $Q \land G$ need not do so. 
Thus the converse of Theorem~\ref{nontermin2} does not hold.
The following example illustrates this point.
\begin{example}\label{double}
{\rm 
Continuing with Example~\ref{double1},
let $Q$ be $X{=}Y$ in the general misbehavior condition.
The implication of the condition is
\[\forall ((X{=}Y \land X{=}s(X1) \land Y{=}s(s(Y1))) \to \]
\[\exists ({\mathit{double(X1,Y1)}}{=}{\mathit{double(X',Y')}} \land X'{=}Y' \land X'{=}s(X1'))).\]
It can be simplified into
\[\forall ((X{=}Y \land X{=}s(X1) \land X1{=}s(Y1)) \to \]
\[\exists (X1{=}X' \land Y1{=}Y' \land X1{=}Y1 \land X1{=}s(X1'))).\]
where $X1{=}s(Y1)$ and $X1{=}Y1$ are in contradiction. Thus the implication does not hold.
However, the goal $\mathit{double}(X,Y) \land X{=}s(X1) \land X{=}Y$  does not terminate.

But there is a more general $Q$ that shows by Theorem~\ref{nontermin3} that the computation for this goal 
either fails or is non-terminating.
Let $Q$ be $X{\geq}Y$.
The implication of the misbehavior condition is
\[\forall ((X{\geq}Y \land X{=}s(X1) \land Y{=}s(s(Y1))) \to \]
\[\exists ({\mathit{double(X1,Y1)}}{=}{\mathit{double(X',Y')}} \land X'{\geq}Y' \land X'{=}s(X1'))).\]
It can be simplified into
\[\forall ((X1{\geq}s(Y1) \land X{=}s(X1) \land Y{=}s(s(Y1))) \to \]
\[\exists (X1{=}X' \land Y1{=}Y' \land X1{\geq}Y1 \land X1{=}s(X1'))).\]
where $X1{\geq}s(Y1)$ implies $X1{\geq}Y1 \land X1{=}s(X1')$. 
The misbehavior condition holds.
So the goal $\mathit{double}(X,Y) \land X{=}s(X1) \land X{\geq}Y$ does not terminate or it fails.
Actually it is non-terminating.
}
\end{example}

\section{Conclusions}

The paper introduced theorems with so-called misbehavior conditions for non-termination and failure as well as termination of linear direct recursive simplification rules in CHR.
Certain goals for a given recursive rule are non-terminating or failing if a certain implication between the built-in constraints of the guard and body of the rule holds. 

We proved a basic theorem for non-termination or failure of the most general goal for recursive rules that consists of their head and guard. A kind of converse also holds: If the misbehavior condition for this goal is violated, it will terminate.
We then gave the main condition for misbeavior. It is parameterised with regard to suitable additional built-in constraints in the goal. 
Finally, a third theorem showed that any goal that contains a misbehaved goal will also be misbehaved. 

\medskip
\noindent{\bf Future Work.}
Having stated the theorems describing non-termination and failure, the immediate next question is how to find the built-in constraints that satisfy the misbehavior condition. This is very likely to be an undecidable problem due to the undecidabilty of termination itself. We can imagine an iterative approach of finding better and better approximations for suitable constraints. 
Another possibility is the systematic enumeration of possible built-in constraints over the involved variables, as one reviewer suggested.

One should extend our approach to a more general class of rules and to other types of recursion.
Our approach readily seems applicable to CHR propagation rules. If other CHR constraints occur in the body of the rule, 
they would have to be abstracted to / approximated by built-in constraints.
Multiple and mutual (indirect) recursion cover the standard formulations of e.g. the Fibonacci and the Ackermann function.
We think that existing rule unfolding techniques for CHR will come handy to replace mutual by direct recursion.

Another open problem is if there is some kind of converse for the main Theorem 2, similar to the one for Theorem 1. A related question is if there are most general built-in constraints for Theorem 2. The answer seems to depend on the expressibility of the built-in constraints in the constraint theory.

Last but not least, it should be investigated how our approach carries over to related languages like constraint logic programming ones and the other rule-based approaches that have been embedded in CHR.
In conclusion, we think this paper can serve as a basic reference and nucleus for a wealth of further research.

\medskip
\noindent{\bf Acknowledgements.} We thank the anonymous referees for their helpful suggestions on how to improve the paper.

\bibliographystyle{abbrv} %!!! alpha is longer
\bibliography{devils,biblio,tfall2005} %,biblio

\begin{thebibliography}{10}

\bibitem{betz_raiser_fru_execution_model_iclp10}
H.~Betz, F.~Raiser, and T.~Fr{\"u}hwirth.
\newblock A complete and terminating execution model for constraint handling
  rules.
\newblock {\em Theory and Practice of Logic Programming}, 10:597--610, 7 2010.

\bibitem{Brockschmidt}
M.~Brockschmidt, T.~Str{\"o}der, C.~Otto, and J.~Giesl.
\newblock Automated detection of non-termination and nullpointerexceptions for
  {J}ava bytecode.
\newblock In {\em Formal Verification of Object-Oriented Software}, pages
  123--141. Springer, 2012.

\bibitem{endrullis2015proving}
J.~Endrullis and H.~Zantema.
\newblock Proving non-termination by finite automata.
\newblock In {\em LIPIcs-Leibniz International Proceedings in Informatics},
  volume~36. Schloss Dagstuhl-Leibniz-Zentrum fuer Informatik, 2015.

\bibitem{fru_chr_book_2009}
T.~Fr{\"u}hwirth.
\newblock {\em Constraint {H}andling {R}ules ({M}onography)}.
\newblock Cambridge University Press, Aug. 2009.

\bibitem{fruhwirth2015constraint}
T.~Fr{\"u}hwirth.
\newblock Constraint handling rules -- what else?
\newblock In {\em Rule Technologies: Foundations, Tools, and Applications},
  pages 13--34. Springer International Publishing, 2015.

\bibitem{fruhwirth2015devil}
T.~Fr{\"u}hwirth.
\newblock A devil's advocate against termination of direct recursion.
\newblock In {\em Proceedings of the 17th International Symposium on Principles
  and Practice of Declarative Programming}, pages 103--113. ACM, 2015.

\bibitem{chrwebsite}
T.~Fr{\"u}hwirth.
\newblock {\em The CHR Web Site -- \tt{www.constraint-handling-rules.org}}.
\newblock Ulm University, 2016.

\bibitem{nonterm2016}
T.~Fr{\"u}hwirth.
\newblock {\em Why Can't You Behave? Non-termination Analysis of Direct
  Recursive Rules with Constraints}, pages 208--222.
\newblock Springer International Publishing, Cham, 2016.

\bibitem{popl2008}
A.~Gupta, T.~A. Henzinger, R.~Majumdar, A.~Rybalchenko, and R.-G. Xu.
\newblock Proving non-termination.
\newblock {\em ACM Sigplan Notices}, 43(1):147--158, 2008.

\bibitem{le2015termination}
T.~C. Le, S.~Qin, and W.-N. Chin.
\newblock Termination and non-termination specification inference.
\newblock In {\em Proceedings of the 36th ACM SIGPLAN Conference on Programming
  Language Design and Implementation}, pages 489--498. ACM, 2015.

\bibitem{kifer}
S.~Liang and M.~Kifer.
\newblock A practical analysis of non-termination in large logic programs.
\newblock {\em Theory and Practice of Logic Programming}, 13(4-5):705--719,
  2013.

\bibitem{payet08}
{\'E}.~Payet.
\newblock Loop detection in term rewriting using the eliminating unfoldings.
\newblock {\em Theoretical Computer Science}, 403(2):307--327, 2008.

\bibitem{mesnardcp}
{\'E}.~Payet and F.~Mesnard.
\newblock A non-termination criterion for binary constraint logic programs.
\newblock {\em Theory and Practice of Logic Programming}, 9(02):145--164, 2009.

\bibitem{mesnardjava}
{\'E}.~Payet, F.~Mesnard, and F.~Spoto.
\newblock Non-termination analysis of {J}ava bytecode, {CoRR} abs/1401.5292,
  2014.

\bibitem{deschreye}
D.~Voets and D.~Schreye.
\newblock A new approach to non-termination analysis of logic programs.
\newblock In {\em Proceedings of the 25th International Conference on Logic
  Programming}, pages 220--234. Springer-Verlag, 2009.

\end{thebibliography}

\appendix
\section{Appendix: Proof of Theorem~\ref{nontermin1}}

Using the proof of Theorem~\ref{nontermin2} as a basis, we now can now prove Theorem~\ref{nontermin1}, too.

\medskip
\noindent {\bf Proof of Theorem~\ref{nontermin1}.}

{\bf The first claim} is that the basic misbehavior condition of this theorem 
implies non-termination or failure of the goal $H \land C$.
This is a simple consequence of Theorem~\ref{nontermin2} where $Q$ is taken to be $\true$.

{\bf The second claim} is that if the basic misbehavior condition 
\[\CT \models \exists (C \land B_{bi})
 \ \land\] 
\[ \ \ \  \ \ \ \forall ((C \land B_{bi}) \to \exists ({B_{ud}}{=}{H'} \land C')).\]
does not hold, then the goal $H \land C$ terminates. 

We prove the claim by case distinction.
If the condition does not hold, then either\\
{\bf Case 1} the existential part of the condition
$\exists (C \land B_{bi})$ is unsatisfiable or\\
{\bf Case 2} the implication part
$\forall ((C \land B_{bi}) \to \exists ({B_{ud}}{=}{H'} \land C'))$ does not hold.

For Case 1, we distinguish two sub-cases;\\
{\bf Case 1.1} $\exists (C \land B_{bi})$ is unsatisfiable, because $\exists C$ is unsatisfiable.\\
{\bf Case 1.2} $\exists (C \land B_{bi})$ is unsatisfiable, but $\exists C$ is satisfiable.\\

{\bf Case 1.1} If $\exists C$ is unsatisfiable, then the state $H \land C$ is failed, so no computation step is possible, the computation trivially terminates and we are done.

{\bf Case 1.2} and {\bf Case 2}
Otherwise $C$ and thus $H \land C$ are satisfiable. Then we can apply the recursive rule to $H$.

In the proof of the base case in Theorem~\ref{nontermin2}, let $Q$ be $\true$.
There we have shown that if the source state $(H \land C \land Q)$ is satisfiable, 
then we can apply the recursive rule and
the resulting target state of the transition is 
$(B'_{bi} \land B'_{ud} \land C \land Q \land H'{=}H \land C')$.

{\bf Case 1.2} Now if $\exists (C \land B_{bi})$ is unsatisfiable, then so is $\exists (C' \land B'_{bi})$ of the target state, since the two formulas are variants. Then the state is failed and we are done.

{\bf Case 2} Otherwise, the implication of the basic misbehavior condition and its variant
$\forall ((C' \land B'_{bi}) \to \exists ({B'_{ud}}{=}{H''} \land C''))$ do not hold.
The applicability condition for applying the recursive rule to $B'_{ud}$ is
$\forall ((B'_{bi} \land C \land H'{=}H \land C') \to \exists ({B'_{ud}}{=}{H''} \land C''))$.
Since by $C' \to \exists (H'{=}H \land C)$ by Lemma~\ref{devils1}, the two implications are equivalent.
Thus the applicability condition does not hold and the recursive rule is not applicable. 
The computation terminates also in this case.

Thus the rule is not applicable and no computation step is possible anymore.
\qed
\medskip

\end{document}